\documentclass[11pt,
prd,aps,amssymb,amsmath  ,tightenlines
]{revtex4}
\usepackage{graphics}
\usepackage[pdftex]{graphicx}

\usepackage{amsfonts}
\usepackage{amssymb}

\newcommand{\ket}{\rangle}
\newcommand{\bra}{\langle}

\begin{document}

\title{ Through the Looking Glass: Bitcoin Treasury Companies}
\author{ B K Meister}
\email{bernhard.k.meister@gmail.com}
\date{\today}
\begin{abstract}
\noindent
Bitcoin treasury companies  have taken  stock markets by storm amassing billions of dollars worth of tokens in hundreds of entities. 
 The paper discusses, how leverage - whether created through corporate debt or investors using stock as loan collateral - fuels this trend.

\noindent
The extension of the binary-choice Kelly criterion to incorporate uncertainty in the form of  the Kullback-Leibler divergence or more generally Bregman divergence is also briefly discussed. 

\noindent

\end{abstract}
\maketitle


\section{Introduction}
\label{sec:1a}
\noindent
The Bitcoin whitepaper\cite{btc} dates back to October 2008, the height of the `great financial crisis' (GFC) and only about a month after the Lehman  bankruptcy. Legacy finance, as it is now sometimes called, was in disarray and 
 a 
 search for alternatives 
 gathered pace. Bitcoin, 
 maybe initially more of a Gedankenexperiment than a practical means of exchange, has over the intervening years grown dramatically. The associated coin BTC  now  has a market capitalisation  of more than two trillion.  
 Thousands of other coins have joined the proverbial gold rush.
For years legacy finance largely ignored these upstarts, but  the growth of BTC's and other tokens' market capitalisations has piqued   curiosity and greed.

\noindent
Companies, and not just individuals, have started hoarding bitcoins.
There are currently over a hundred companies 
that are  devoted to this one task - the holding of crypto-assets. Many market themselves as vehicles whose mission it is to acquire ever-more and ever-faster crypto-tokens financed by a mixture of equity and various form of debt issuance. 
Puzzlingly, their market capitalisation can be significantly higher than the  net asset value of their crypto-assets. This is especially surprising when the non-crypto business of these   companies is often between insignificant and non-existent. How can this be? The paper suggests possible  answers, which go  beyond irrational exuberance. 

\noindent
In the next section popular new performance indicators  are briefly introduced and in the subsequent sections the role of leverage is explored for both investors and the treasury companies themselves.
The penultimate section of the paper is devoted to an extension of the one-dimensional Kelly criterion to cases where the winning probability is uncertain. 
A conclusion rounds of the paper.

\section{ Key performance indicators `reimagined' }
\noindent 
One of the key performance indicators (KPIs) touted by  crypto-treasury companies is mNAV, which is the ratio of enterprise value\footnote{In the case of an `idealised' company, which is the only company we consider in this section, the enterprise value is equivalent to the market capitalisation. We assume the company is debt free, no preferred equity has been issued and (for completeness sake) no pension liabilities exist.} to the value of the crypto-tokens held. 
If this is above one issuing equity to purchase further tokens is  value  accretive. To see this take a simple example. Assume the token trades at $\$1$, and the mNAV is $4$. Further assume  $4$ shares have been issued trading at $\$1$ each, and one token is in the companies treasury.
Next a share is issued at the prevailing  price  of $\$1$ and one token is bought.
Assuming an unchanged mNAV, the   price per share becomes $\$ \, 8/5$ an increase of $80\%$.
Magically, everyone seems to be better off. How can this be? Is it due to  a suspension of disbelief, since the gap between net asset value (NAV) and the market capitalisation  has widened\footnote{There are also tax and other hurdles that prevent investors from buying tokens directly. 
For example, in Japan holding Bitcoin can be less tax efficient than stock ownership. Cryptocurrency capital gains tax in Japan  is up to $55\%$, while the comparable tax on stocks is only about $20\%$. These tax rates are currently under review.}. 
Why overpay for fractional ownership of crypto-tokens via an intermediary, when either through direct ownership  or some other intermediary, as should be possible in a competitive market,    the same exposure can be obtained cheaper?

\noindent
`Bitcoin per share', the fraction of Bitcoin held per share, is also a popular KPI. A rise `Bitcoin per share'  is viewed as beneficial for the shareholders. This can  be achieved with debt  or with equity issuance, as we saw above,  for mNAV larger than one. Debt  issuance is interesting, since it enables the shift of risk to outside parties. This is also true for  borrowing by shareholders against portfolio value. The paper considers in the next section credit transfers by borrowing of shareholders but not by the company itself. The different types of credit transfers are not interchangeable, even if in both cases risk is shifted away from the `centre'.

\noindent
Other KPIs including BTC yield, BTC gain and BTC \$ gain have also been bandied about but are maybe not 
as helpful, or at least add little beyond the metrics mentioned above, to understand the price behaviour. Once mNAV approaches one,   other KPIs as well as price impact\footnote{For an analysis of strategies incorporating price impact see \cite{bkm2016}. Here we are interested with the price impact on Bitcoin  by companies buying a substantial amount of the outstanding tokens.}  gain in importance.  
In the next section we return to mNAV.
\section{Mirror images: Is collaterlized borrowing a  virtuous, a vicious or  a virtual cycle? }
\noindent
In   section II of a paper on `rational market failures' \cite{meister2022}
a cyclical strategy was described that allowed risk to be passed from stock investors to credit providers. With sufficient market impact an investor 
was able to extract money. Next, an outline of how the process works. The investor would buy a particular stock, drive up the stock price, since one-sided demand can lift the price, and the portfolio value. 
These portfolio gains the investor could use as collateral to  borrow against. This would lead to a virtuous cycle for the stockholders, 
and a potentially vicious cycle for the  lenders, 
which would accumulate risk. In extreme cases the investor could extract cash and come out ahead, even if the stock eventually crashes.  For details, see the earlier paper.

\noindent
The `innovation' of crypto-treasury companies is to exploit  the counter-intuitive  situation of having a mNAV larger than one. In its simplest form the issuance of stock converted back into crypto-tokens raises the stock price, followed by more stock issuance. Additional stock in one scenario is bought by existing holders of the stock, who borrow against the ever increasing value of their stock portfolio to finance the new purchases. This is true even without  any  overt price impact on crypto-tokens. It wouldn't be surprising if the steady buying by treasury companies of an assets that is in the case of Bitcoin of fixed supply will in addition raise the price of the token. The efficiency and thereby viability of the equity-issuance  strategy only depends on banks or other credit providers treating stock of  Bitcoin  treasury companies  like other assets that can be deposited as collateral for loans with a reasonable haircut. When collateral requirements are  raised,  the attractiveness of these companies as endlessly equity-issuing financial perpetuum mobiles may disappear suddenly. 
Next a look at companies instead of investors raising leverage through debt.

\section{
Is Debt the Fulcrum of Finance Treasury Companies? }
\noindent
This section  considers a Bitcoin treasury company raising money through a zero coupon bond to add to an existing  trove of tokens. This is a simplification of what one observes in markets, since treasury companies are often serial issuers with an eclectic mix of offerings.   Nevertheless, this pared down model provides  insights. Next,  the relevant quantities are defined. 
The Bitcoin price is $B_0$ at time $t=0$ and described by a geometric Brownian motion with volatility $\sigma$ and drift $\mu$,
\begin{eqnarray}
dB_t = \mu B_t dt + \sigma B_t dW.\nonumber
\end{eqnarray}

\noindent 
The quantities  $A  \, \& \,\, Y$ are denominated in $[ BTC ]$, 
$C_{BS}\,\, \&\,\, P_{BS}$ in $ [\$/BTC  ]$\footnote{Option prices are often quoted dollars, and the fact that the price is per contract (or certain number of shares) is implicit.}, 
$X$ in $ [\$ ]$, 
$T$ in $ [ 'time'  ]$,  $B_t $ in $ [  \$ /BTC  ]$,  
$r\,\, \& \,\,R$ in $ [ 1/'time'  ]$, $\sigma^2$ in $ [ 'time'  ]$ and
 $Q$ is a dimensionless number.

\noindent 
 Pre-issuance the amount of Bitcoin held in the company is $A$. At time  $t=0$, $Q$ units of a zero-coupon bond are issued at price $X$ per  unit and are to be redeemed by the company for $1\$$ per unit at time $T$.
 
\noindent 
The issuance proceeds of $X Q$ are immediately converted into $ XQ/B_0$ Bitcoins.
The company holds a fixed amount of Bitcoins, i.e. $A+  XQ/B_0:=Y$,  until the bonds mature. Bankruptcy can only occur at $T$.
At $T$ the following happens, the bond investors receive
\begin{eqnarray}
 Min  \Big[ 1\$\,Q  ,  B_T Y   \Big]\nonumber
\end{eqnarray}
and the residual or unwind value of the company is
\begin{eqnarray}
 Max  \Big[  B_T Y- 1\$\,Q\, ,0  \Big].\nonumber
\end{eqnarray}
We take $B_0\,\,\,\&\, A$ to be inputs, whereas the amount $Q$ is to be optimised to maximise the post-issuance share price, i.e. just after $t=0$.

\noindent 
What is the value of the stock  post debt issuance? Since the company has an obligation to service its the debt,  the stock is an option on Bitcoin. The value at time $t=0$ is given by $Y$ units of the Black-Scholes formula 
for   an European call
\begin{eqnarray}
 Y\,\,C_{BS} (  B_0 ,  1\$ \,Q/Y,\sigma,R,r,T), \nonumber
\end{eqnarray}
with $B_0$ the price of the underlying, $ 1\$ \, Q/Y$ the strike\footnote{Here and elsewhere $1\$ $ is added,  
 to have dimensional consistency.
 }, $\sigma$ the Bitcoin implied volatility, $T$ the maturity, and for completness $R\,\, \&\,\, r$ are the relevant interest rates.  The Bitcoin yield curve was modelled theoretically in \cite{bhm} and empirically  in \cite{new} provides $r$, while   $R$ is the relevant interest rate for   the riskless USD curve.  The Bitcoin interest rate applied in the Black-Scholes formula can   be modified by storage cost and convenience yield. 
 For  short maturities the price can also be interpolated  from   data found on Deribit or other exchanges.  
 If the Bitcoin to dollar exchange rate drops below $Y/(1\$Q)\geq B_T$ the option expires worthless and the company retains zero value. In this, like the rest of the analysis, bankruptcy cost is ignored.

 \noindent
 Similarly, one can also determine the value of $X$, since the package of an appropriately structured put and the risky issued bond  is a risk-free bond. In more detail, $XQ$ is the difference between the risk-less price, which only depends on discounting, i.e.  $1\$ \,e^{-RT}Q$, and an put embedded in the bond, since the $Y$ Bitcoins held by the company will be insufficient the repay the obligations at $T$, if the Bitcoin prices falls below $B_T\leq 1\$ Q/Y$.  
 $X$ is calculated  using again the   Black-Scholes model, or based on prices observed on the cryptocurrency options exchange Deribit or similar venues.  
 The    expression  $( 1\$ \,e^{-RT}-X)Q$ is   equivalent to the Black-Scholes put price for $Y$ units of $P_{BS} (  B_0 , 1\$ \,  Q/Y,\sigma,R,r,T )$, since the put payout has the form
 \begin{eqnarray}
 Max  \Big[  1\$\,Q\, -B_T Y,0  \Big],\nonumber
\end{eqnarray}
and compensates  for the shortfall the bond investors suffer, if the company goes bankrupt.
 Put and call prices are linked by put-call parity, i.e.  $Y C_{BS} (B_0 ,  1\$ \, Q/Y,\sigma,R,r,T )-Y P_{BS} (  B_0 ,  1\$ \, Q/Y,\sigma,R,r,T )= YB_0e^{-rT}-1\$ \,Qe^{-RT}$.

\noindent 
 As stated above, one can view the share of the treasury company for $0<t\leq T$ as an option on the value of Bitcoin.  If  $Y \, C_{BS} (B_0,1\$\,Q/Y,\sigma,R,r,T )$ for a particular choice of $Q$, or alternatively the price for this structure on Deribit, is larger than $B_0 A$, then issuing bonds creates instantaneous value for shareholders, and an instability exists until the issuance feedback loop is exhausted. If this is possible in theory is discussed in the next paragraph. If it is exploitable in  practice depends on the prices quoted on Deribit, liquidity and financing constraints. Since this question is of a more applied nature\footnote{Complications can for example occur if as part of an  initial issuance the company agrees to  limits  future issuance.}, it will be explored elsewhere.

 \noindent 
 Can issuance of debt  increase  shareholder value in the Black-Scholes framework? This question is  equivalent  to finding a value $Q$ for which the expression
 \begin{eqnarray}
   Y \, C_{BS} (B_0,1\$ \,  Q/Y,\sigma,R,r,T )- B_0 A \nonumber
\end{eqnarray}
is positive.
First replace, using the put-call parity, $C_{BS} (B_0,1\$ \,  Q/Y,\sigma,R,r,T )$
to get 
\begin{eqnarray}
  Y \,P_{BS} (  B_0 ,  1\$ \, Q/Y,\sigma,R,r,T )+ YB_0e^{-rT}-1\$ \,Qe^{-RT} - B_0 A.\nonumber
  \end{eqnarray}
This is equivalent to 
\begin{eqnarray}
 B_0Y(e^{-r T}-1)=(B_0A+X Q )(e^{-r T}-1),\nonumber
  \end{eqnarray}
 which is equal to zero,
for $r=0$,  negative for $r>0$, and   positive   $r<0$.  
Therefore, zero-coupon bond issuance is  for $r< 0$   accretive for stockholders.
  At what stage the stock market recognises the gain, i.e. pre- or post-announcement, or maybe post-approval or  -issuance, is debatable.

  \noindent 
  Next, the analysis is extended to two issuances. 
  We assume debt is issued    in two tranches at $t=0$ with the amounts $Q_1$ and $Q_2$, and either priced sequentially with the existence of the second tranche   not known when $X_1$, the fair value for the bond, is determined, or priced as one tranche of size $Q_1+Q_2$.
 
 \noindent 
Three cases are considered, for each there are three constraints and one inequality. The cases are named A,B \& C, where A corresponds to one issuance of $Q_1$ units of debt, B corresponds to  the issuance of $Q_1$ and $Q_2$ units of debt in sequence, but almost simultaneously,
  and C corresponds to one issuance of $Q_1+Q_2$ units of debt.

\noindent 
 I. The amount converted to Bitcoin is in the three scenarios
 
\begin{eqnarray}
 A: \,\,\,\,&&AB_0+X_1Q_1=Y_1 B_0 ,\nonumber\\
 B: \,\,\,\,&&AB_0+X_1Q_1+ X_2Q_2=Y_2 B_0 ,\nonumber\\
 C: \,\,\,\,&& AB_0+\overline{X}( Q_1+ Q_2)=\overline{Y}  B_0.\nonumber
\end{eqnarray}

\noindent 
II.The value of Put is in the three scenarios

\begin{eqnarray}
  A: \,\,\,\,&&(e^{-RT}1\$\,-X_1)Q_1=Y_1 P\Big(B_0,1\$\,\frac{Q_1}{Y_1}, \sigma,R-r,T\Big) ,\nonumber\\
 B: \,\,\,\,&&(e^{-RT}1\$\,-X_2)Q_2=\frac{Q_2}{Q_1+Q_2}Y_2 P\Big(B_0,1\$\,\frac{Q_1+Q_2}{Y_2}, \sigma,R-r,T\Big) ,\nonumber\\
 C: \,\,\,\,&& (e^{-RT}1\$\,-\overline{X})(Q_1+Q_2)=\overline{Y} P\Big(B_0,1\$\,\frac{Q_1+Q_2}{\overline{Y}}, \sigma,R-r,T\Big).\nonumber
\end{eqnarray}

\noindent 
III. Put-Call-Parity is in the three scenarios

\begin{eqnarray}
 A: \,\,\,\, && Y_1 C\Big(B_0,\frac{1\$\,Q_1}{Y_1}, \sigma,r,T\Big)- Y_1 P\Big(B_0,\frac{1\$\,Q_1}{Y_1}, \sigma,R,r,T\Big)= Y_1 B_0e^{-rT}-e^{-RT} 1\$\,Q_1 ,\nonumber\\
 B: \,\,\,\,&&Y_2 \Big(C\Big(B_0,1\$\,\frac{ Q_1+ Q_2}{Y_1}, \sigma,R,r,T\Big)-  P\Big(B_0,1\$\,\frac{Q_1+Q_2}{B_1}, \sigma,R,r,T\Big)\Big) \nonumber\\
 &&= Y_2 B_0e^{-rT}-e^{-RT} 1\$\,( Q_1+ Q_2) ,\nonumber\\
 C: \,\,\,\,&&\overline{Y}\Big(C\Big(B_0,1\$\,\frac{ Q_1+ Q_2}{\overline{Y}}, \sigma,R,r,T\Big)- P\Big(B_0,1\$\,\frac{ Q_1+ Q_2}{\overline{Y}}, \sigma,R,r,T\Big)\Big) \nonumber\\
 &&=\overline{Y} B_0e^{-rT}- e^{-RT}1\$\,( Q_1+ Q_2).\nonumber
\end{eqnarray}

\noindent 
IV. The Stock Value adjustment is therefore   in the three scenarios

\begin{eqnarray}
  A: \,\,\,\,&&Y_1 C\Big(B_0,1\$\,\frac{Q_1}{Y_1}, \sigma,R,r,T\Big) - AB_0,\nonumber\\
 B: \,\,\,\,&&Y_2 C\Big(B_0,1\$\,\frac{ \,Q_1+\,Q_2}{Y_2}, \sigma,R,r,T\Big) -AB_0 ,\nonumber\\
 C: \,\,\,\,&& \overline{Y}   C\Big(B_0,1\$\,\frac{ Q_1+Q_2}{\overline{Y}}, \sigma,R,r,T\Big)-AB_0.\nonumber
\end{eqnarray}

\noindent 
The first and third case are evaluated identically, and  have been dealt with above, whilst for the second case  the stock value change is  
\begin{eqnarray}
Q_1( X_1-X_2)
 ,\nonumber
\end{eqnarray}
for $r=0$, 
and issuance of a zero coupon  can at best be value neutral and otherwise  destroys 
value. Next, $r\leq 0$ is considered.
\noindent 
  The stock value adjustment has  now the form
 \begin{eqnarray}
&& (AB_0+Q_2 X_2) (e^{-rT}-1)+Q_1( e^{-rT}X_1-X_2),\nonumber
\end{eqnarray}
which is positive for 
 \begin{eqnarray}
&& - \frac{1}{T}\log\Big(\frac{(Q_1+Q_2)X_2+AB_0}{AB_0+Q_2 X_2+Q_1X_1}\Big)> r.\nonumber
\end{eqnarray}
For   small $r $ one has the expansion
 \begin{eqnarray}
&&Q_1(  X_1-X_2) -rT(AB_0+Q_2 X_2+  Q_1X_1)+ O(r^2).\nonumber
\end{eqnarray}
\noindent
Next a section on the asymmetric influence of uncertainty on optimal investment fractions.
  \section{The Kelly criterion in uncertain markets  }
\noindent
 The one-dimensional Kelly criterion is in this section extended to  uncertain winning probabilities.
The   Kullback-Leibler (KL) distance or some more general Bregman divergence between the assumed and real probabilities captures the idea of lack of knowledge. 
The application of the KL distance to the Kelly criterion was already discussed in \cite{meister2024} but in the earlier paper the emphasis was on large deviations and the ability to distinguish strategies associated with different investment fractions.

\noindent
Before a toy model for uncertainty and the influence on the investment fraction is discussed, 
 we study the deliberate over- or under-investment in a situation with a known winning probability. It was  shown in \cite{meister2011} how  this affects  the portfolio growth rate negatively as well as asymmetrically. This result can be understood intuitively, since if one underinvests, the growth underperforms but is bounded by zero. If one instead over-invests,  the growth is not bounded by zero and instead can turn negative. 
Next, the intuition is confirmed by expanding the growth rate in terms of a small parameter linked to the optimal fraction. 
If the winning probability is known to be $p$ in the standard double-or-nothing game, and one varies the optimal investment fraction by an infinitesimal amount $\epsilon$, i.e. $f= 2p-1 +\epsilon$, then this leads to 
\begin{eqnarray}
 && p \log ( 1+ f) + (1-p)\log(1-f ) = p \log(2p +\epsilon) +(1-p) \log(2-2p -\epsilon) \nonumber\\
 &=&H(p) +1 + p \log\Big(1+\frac{\epsilon}{2p}\Big)+(1-p) \log\Big(1-\frac{\epsilon}{2-2p}\Big)
 \nonumber\\
 &=&H(p) +1 + p \Big(\frac{\epsilon}{2p}-\frac{\epsilon^2}{8p^2}+\frac{\epsilon^3}{24p^3}\Big)-(1-p) \Big(\frac{\epsilon}{2-2p}+\frac{\epsilon^2}{8(1-p)^2}+\frac{\epsilon^3}{24(1-p)^3}\Big)+O\big(\epsilon^4\big)
 \nonumber\\
 &=&
H(p) +1- \frac{1}{8(1-p)p}\epsilon^2-\frac{2p-1}{24 (1-p)^2p^2}\epsilon^3+O\big(\epsilon^4\big).\nonumber
 \end{eqnarray}
  Let us phrase the above result in terms of  the existence and sign of the quadratic and cubic terms in $\epsilon$. The growth is suboptimal for non-zero $\epsilon$, i.e. the leading term is quadratic and has a negative sign, and the loss in the portfolio growth rate due to over-investment is larger than due to under-investment for the same absolute value in $\epsilon$, i.e. the cubic term has a negative sign.






\noindent
The inputs of the toy model for investment uncertainty discussed next are the value of the KL divergence $D_{\mathrm{KL}} \left( q  \parallel p  \right)$ set equal to $\alpha$, which is  assumed to be small, i.e. $\alpha \ll 1$, as well as the  probability $q$, which is assumed to be larger than $1/2$. The  separation of $q$ from the unknown  `correct' probability is given by the KL divergence $\alpha$. To avoid    possible solutions for $p$ from falling below $1/2$, which would cause the associated investment fraction to be negative, $q$ is set sufficiently above $1/2$. We  further define  $p=q+ \epsilon$, where $\epsilon$ is small. 
An expansion of $D_{\mathrm{KL}} \left( q  \parallel p  \right)$ in terms of $\epsilon$ has the form  
 \begin{eqnarray}
 &&D_{\mathrm{KL}} \left( q  \parallel p  \right)= p \log\Big(\frac{p}{q}\Big) +(1-p)  \log\Big(\frac{1-p}{1-q}\Big)\nonumber\\
  &=& \sum_{n=1}^{\infty} \Big(q+\epsilon  \Big) (-1)^{n+1}\frac{ \epsilon^n}{n  \,  \, q^n  }     -\Big(1 -q-\epsilon  \Big)\frac{\epsilon^n }{n \, \,(1-q)^n }    \nonumber\\
&=&  \sum_{n=2}^{\infty}\frac{ \epsilon^n}{ n^2 -n }\Big( \frac{ (-1)^{n+1 }}{ q^{n-1}  }     -  \frac{1 }{(1-q)^{n-1}   } \Big) = \sum_{n=2}^{\infty}\frac{ \epsilon^n}{ n^2 -n }\Big( \frac{ (-1)^{n +1}(1-q)^{n-1} -q^{n-1}}{ q^{n-1} (1-q)^{n-1}  }      \Big) \nonumber\\
&=&   \frac{1}{2(1-q)q}\epsilon^2  -\frac{1}{6}  \frac{(1-q)^2-q^2}{(1-q)^2q^2} \epsilon^3+  \frac{1}{12}  \frac{(1-q)^3+q^3}{(1-q)^3q^3} \epsilon^4-\frac{1}{20}  \frac{(1-q)^4-q^4}{(1-q)^4q^4} \epsilon^5+O(\epsilon^6),\nonumber\\
&=&   \frac{1}{2(1-q)q}\epsilon^2  +\frac{1}{6}  \frac{2q-1}{(1-q)^2q^2} \epsilon^3+O(\epsilon^4).\nonumber
 \end{eqnarray}
 If, as stated above, both $D_{\mathrm{KL}} \left( q  \parallel p  \right)$, equal to $\alpha$, and $q$ are fixed, then in lowest approximation,   
as one solves the equation up to a quadratic terms, one gets two solutions for $\epsilon$.
One can extend the analysis to the cubic term without altering the result\footnote{If one includes the cubic term and assumes reasonable values for $\alpha$ and $q$, one gets also in the relevant range two solutions $p_1 \, \& \,p_2$  with $p_1\leq q\leq p_2$, then $ |q-p_1|\geq |p_2-q|$ such that $p_1 + p_2 \leq 2q$. This would mean that up to cubic terms the average of the two individually `optimal' fractions would be slightly lower than  $q$. As a consequence, the optimal fraction is lower, i.e. $f=(2q-1) (1-\frac{ 2}{ 3}\alpha) +O(\alpha^2)$,  and also separately one can show that the growth rate is reduced  by series in $\alpha$, i.e. $\alpha- \alpha^2 \frac{(2q-1)^2}{18 q(1-q)}+ O(\alpha^3)$. }.
 Various cases can be considered. One can consider   the worst case, i.e. the minimal solution for $p$, the best case, i.e. the maximal solution for $p$, and a `least informative' case, i.e. with equal weights for both possible solutions of $p$. 
In the equal weighted case, 
due to the asymmetry associated with over- and under-investment, as described above,   the optimal investment fraction is lower than $2q-1$. 
To summarise, uncertainty about the right value for the winning probability leads to a lowering of the growth rate and the optimal fraction.  An attractive  formula for the adjusted investment fraction that captures this heuristically is $ f^* \exp(- \lambda
\alpha )$, where $f^*$ is the optimal uncertainty-free fraction $2q-1$, and $\lambda$ is a risk-aversion parameter. 


\noindent
This result can be extended from KL divergence to the more general Bregman divergence   by doing the relevant Taylor expansions for cases of interest like the Itakura–Saito distance. A similar `conservatism' can be observed.

\noindent
The rationale for the fractional Kelly criterion, where one invests a fraction of the `optimal' investment amount, can also be reevaluated in the light of the above result.  A more conservative invest fractional investment is not due to excessive risk aversion, but instead due to uncertainty about the state of the world, i.e. in the simplest case the winning probability of a coin.

 \noindent
The conclusion attempts again to shed light on treasury  companies. 
 
\section{Conclusion } 
\noindent

\noindent
Crypto treasury companies with a rising `Bitcoin per share' and mNAV above one seem
to have discovered a financial perpetuum mobile and are  eager to recruit  followers, since copy-cats compete for the  restricted  supply of Bitcoins and in the process  push up prices.
Is there a fault in their logic? First, credit risk aggregates   in the system, when these stocks are used as collateral to obtain leverage to race ahead in the accumulation process. This  risk has to be managed to mitigate potential losses, when there is a dramatic price reversal of BTC   or mNAV sufficiently compresses leading to portfolio liquidations. Second, if price impact of buying limited supply tokens is central to the scheme, then there is the  possibility  of a widening gap between the price and the value of these tokens. As Fischer Black once stated for equities, the gap between price and value in an `efficient market'\footnote{Fischer Black in ``Noise", Journal of Finance, vol. 41, pp. 529–543 (1986) stated that he thought `markets are efficient almost all of the time. ``Almost all'' means $90\%$ of the time'. This statement was and to a lesser extent still is controversial.} is `almost all' of the time not  larger than $100\%$. Constraints on value exists even for crypto tokens, since a market capitalisation much beyond the current single digit trillion dollar amount  has a good chance to unbalance not just the  financial system and the economy in general, but also politics.   

\noindent
Can the gap between market capitalisation and net asset value hold indefinitely as more equity is issued, bought by current holders, and used as collateral for loans to buy further tokens?  
This seems unlikely, since the strategy has a weak point: Credit providers.
The willingness of credit providers to lend against stocks is surely limited, since their risk appetite is bounded. What can credit providers do?  They should be cognizant of the gap between market capitalisation and  NAV and align their lending to the limited liquidation value under market stress  of large crypto-token portfolios. It seems  brokers are aware of the risk and are restricting lending against concentrated positions in token treasury companies\footnote{This is suggested by Reddit discussion forums and similar anecdotal evidence.}. This could  break the cycle, but, as the saying goes, `the market can stay irrational longer than investors can stay solvent'.

\noindent
This weakness is less evident for the strategy described in section IV based on debt issuance, which  
aims to   
generate instantaneous value for shareholders. 
Let's recap: The idealised Bitcoin treasury company, which at the start solely holds crypto currency, issues a zero-coupon bond. The money raised is used to buy further Bitcoins. The bond is not risk-free, since  Bitcoin falling below a threshold vis-a-vis  dollar at bond maturity leads to bankruptcy. A riskless zero-coupon investment is the combination of the bond issued by the Bitcoin treasury company and  a put, which indemnifies bond investors against losses in case of default. 
 
 \noindent
It was shown that the company value remains unaffected by an individual zero-coupon debt issuance of any size or maturity, independent of Bitcoin volatility, USD interest rate, and initial Bitcoins held by the company, if the Bitcoin interest rate is set to zero. Whilst issuance is accretive, if $r$ is negative.  
A  more thorough  investigation  requires the study of more elaborate hybrid, i.e. equity plus debt, funding mechanisms which treasury companies are now employing. The analysis also ignored that one-sided  activity leads to buying pressure on Bitcoin\footnote{
%
Unlike the temporary price pressure from an index inclusion of an equity - which doesn't normally affect fundamental value - this effect should be more pronounced for harder to evaluate cryptocurrencies, since    the line between transitory  and  permanent market impact is  more easily blurred.
}, and can, through `price-impact', create   market momentum. 
This effect and the resulting dynamic will be studied  elsewhere together with  general constructions involving  leveraged positions in     liquid assets. 


 \noindent
The history of finance is full of wondrous schemes.  After the South Sea bubble even Newton was supposed to  have said, `I can calculate the motion of heavenly bodies, but not the madness of people', and similar events in France, where the `Mississippi Bubble'  ended badly for the buccaneer  
John Law, undermined the established order, and 
prepared the ground for the coming revolution.

\noindent
In the second part of the paper,  simple information theoretic ideas 
are   employed  to understand the influence of uncertainty on decision making, and an asymmetric effect on the optimal fraction for the binary-choice Kelly criterion    was   observed as the winning probability becomes uncertain. 

\noindent
 The paradox of high mNAV crypto treasury companies feverishly issuing equity is captured by the White Queen  in Lewis Carroll's `Through the Looking Glass', when she  says, `[t]he rule is, jam to-morrow and jam yesterday – but never jam to-day'.
 
\end{document}